\documentstyle[12pt]{article}

\begin{document}
\centerline{{\bf On neutron star/supernova remnant associations}}
\vskip 5mm
\centerline{{\bf V.V.Gvaramadze}\footnote{{\it Address for correspondence}:
Krasin str. 19, ap. 81, Moscow, 123056, Russia; e-mail:
vgvaram@mx.iki.rssi.ru}}
\vskip 5mm
\centerline{Abastumani Astrophysical Observatory, Georgian Academy of
Sciences,}
\centerline{A.Kazbegi ave. 2-a, Tbilisi, 380060, Georgia;}
\centerline{Sternberg State Astronomical Institute, Universitetskij
Prospect 13,}
\centerline{Moscow, 119899, Russia;}
\centerline{Abdus Salam International Centre for Theoretical Physics,}
\centerline{Strada Costiera 11, P.O. Box 586, 34100 Trieste, Italy}
\newpage

\centerline{{\bf Abstract}}
\vskip 5mm
It is pointed out that a cavity supernova (SN) explosion of a {\it moving}
massive star
could result in a significant offset of the neutron star
(NS) birth-place from the geometrical centre
of the supernova remnant (SNR). Therefore: a) the high implied
transverse velocities of a number of NSs (e.g.  PSR B\,1610-50,
PSR B\,1757-24, SGR\,0525-66) could be reduced;
b) the proper motion vector of a NS should not necessarily
point away from the geometrical centre of the associated
SNR; c) the circle of possible NS/SNR
associations could be enlarged. An observational test is discussed, which
could allow
to find the true birth-places of NSs associated with
middle-aged SNRs, and thereby to get more reliable
estimates of their transverse velocities.

\section{Introduction}

Though the number of secure associations between NSs
and SNRs continues to grow (Caraveo 1993, 1995;
Allakhverdiyev et al. 1995,
Kaspi 1996, 1998, 2000; Frail 1998; Helfand 1998, Mereghetti 1998, 1999;
Marsden et al. 1999), it is considered that many of claimed associations are
spurious
and are merely the results of superposition (e.g. Gaensler \& Johnston
1995a,b; cf. Lorimer, Lyne \& Camilo 1998). It were proposed five criteria
for the evaluation of possible NS/SNR associations which come to the following
questions (Kaspi 1996):\\
\hskip 5mm
-- do independent distance estimates agree?\\
\hskip 5mm
-- do independent age estimates agree?\\
\hskip 5mm
-- is the implied transverse velocity reasonable?\\
\hskip 5mm
-- is there evidense for any interaction between the NS and SNR?\\
\hskip 5mm
-- does the proper motion vector of the NS point away from the SNR centre?\\
The last question is considered the most important one since
"a proper motion measurement has the potential to disprove an
association regardless of the answers to the other questions" (Kaspi
1996).

In this Letter we point out that a cavity SN explosion of a
{\it moving} massive star could result in a significant offset of the NS
birth-place from the geometrical centre of
the SNR (Sect. 2). Three important consequences can be drawn from this:
1) the implied transverse velocity of the NS
(i.e. the velocity derived from the displacement of the NS from
the geometrical centre of the SNR) could be significantly overestimated;
2) the proper motion vector of the NS should not necessarily point
away from the geometrical centre of the associated SNR (it even could
be directed to the centre of the SNR!); 3) the circle of possible
NS/SNR associations could
be enlarged (Sect. 3). These facts are quite obvious, but have been largely
overlooked in studies of NS/SNR associations.
It is also suggested that the birth-place of a NS could be marked by a
nebula of thermal X-ray emission (Sect. 3).
The possible detection of such nebulae
will allow to get more reliable estimates of transverse velocities of NSs
associated with middle-aged SNRs.

\section{Off-centred cavity SN explosion}

Massive stars (the progenitors of most of SN stars; e.g. van den Berg \&
Tammann 1991; Tammann, L\"offler \& Schr\"oder 1994) strongly modify,
during their evolution from the main-sequence (MS) to the SN explosion, the
ambient medium by ionizing emission and winds, that results in the origin
of a system of cavities and shells (Avedisova 1972; Dyson \& de Vries
1972; Dyson 1975; Castor, McCray \& Weaver 1975; Steigmann, Strittmatter
\& Williams 1975; Weaver et al. 1977; McCray 1983; McKee, Van Buren \&
Lazareff 1984; D'Ercole 1992).
The subsequent interaction of the SN blast wave with the
reprocessed circumstellar and interstellar medium results in the
origin of a SNR (e.g. Fabian, Brinkmann \& Stewart 1983; Shull et al. 1985;
McKee 1988; Ciotti \& D'Ercole 1989; Chevalier \& Liang 1989; Chevalier \&
Emmering 1989; Franco et al. 1991; McCray 1993; Brighenti \& D'Ercole
1994). The structure of a young ($\leq 10^3$ years)
SNR is mostly determined by the interaction of
the SN blast wave with circumstellar structures created during the late
evolutionary stages of the SN progenitor star (e.g. McCray 1993;
Garsia-Segura, Langer \& Mac Low 1996; Borkowski
et al. 1996), while the appearance of a middle-aged SNR could be
affected by the interaction of the SN blast wave with large-scale
structures created in the interstellar medium by the stellar ionizing
emission (the shell of neutral gas around a Str\"omgren sphere; e.g.
Shull et al. 1985) and/or  (fast) stellar wind (the density jump at the edge
of a stalled wind-driven bubble or the dense large-scale shell
swept-up from the
interstellar medium by an expanding bubble; e.g. Ciotti \& D'Ercole 1989;
Franco et al. 1991; D'Ercole 1992; Gvaramadze 1999a,b).

It is clear that the SN explodes in the centre of the system of cavities
and shells if the space velocity of the SN progenitor star is equal to zero.
In this case the birth-place of the SN stellar remnant (e.g. a NS) coincides
with the geometrical centre of the future SNR, and therefore the implied
transverse velocity of the stellar remnant is equal to the true one.

But the SN explosion site could be significantly offset
from the geometrical centre of large-scale structures created in the
reprocessed ambient medium if the
massive star moves relative to the interstellar medium (it is known that
most of massive stars have a space velocity of a few
${\rm km}\,{\rm s}^{-1}$; e.g. Vanbeveren, De Loore \& Van
Rensbergen 1998). E.g. a 25 $M_{\odot}$
star moving with the velocity of 2 ${\rm km}\,{\rm s}^{-1}$ travels from the
centre of the MS bubble for about 18 parsecs. The stellar
motion does not affect the spherical shape of the bubble (here and below
we assume that there is no large-scale density inhomogeneities in the ambient
interstellar medium) since the sound speed in the
hot interior of the bubble is about two orders of magnitude larger than
the velocity of the star (e.g. Weaver et al. 1977). Correspondingly, the
(middle-aged) SNR also acquires the spherical shape even if the SN
exploded far from the centre of the wind-driven bubble (e.g.
R\'{o}\.{z}ychka et al. 1993). The off-centred cavity SN explosion
results, however, in the inhomogeneous distribution of the surface
brightness over the SNR's shell, that could explain the arc-like
appearance of some of middle-aged SNRs (see R\'{o}\.{z}ychka et al. 1993,
Brighenti \& D'Ercole 1994). Note that the standard
explanation of the origin of incomplete shells implies the interaction of
the (Sedov-Taylor) blast wave with the inhomogeneous (e.g. cloudy)
interstellar medium. Therefore, the non-detection of a dense large-scale
cloud nearby to the arc-like SNR could serve as an indirect evidence that
this SNR is generated by a moving massive star (cf. Brighenti \& D'Ercole
1994). The anonymous referee mentioned that "the direction of motion of
the NS ought to depend on its location relative to the bright part of the
shell". This is correct, however, only in the absence of other factors
affecting the brightness distribution over the SNR's shell. The existence
of large-scale density gradients in the interstellar medium
and/or magnetized wind-driven shells leads to the more complex appearance
of SNRs (see e.g. Gvaramadze 1999b), and therefore the enhanced brightness
of a part of the shell not necessarily ought to be due
to the proximity of the SN explosion site.
The detailed study of this important issue is beyond the scope of
this Letter and will be carried out elsewhere.

It should be mentioned that only a fraction of middle-aged and old SNRs
is the result of cavity SN explosions. Indeed, one can show
(e.g. Brighenti \& D'Ercole 1994) that only slowly moving ($1-2 \, {\rm
km}\,{\rm s}^{-1}$) and/or very massive stars explode inside the
wind-driven bubbles created during the MS stage of their
evolution. But even if a massive star is fast enough to cross the stalled
MS bubble it could again find itself in the bubble interior if
it ends its evolution as a red supergiant (RSG) star
(i.e. if the zero age main sequence mass of the
star is $\leq 15-20 \, {\rm M}_{\odot}$; e.g. Vanbeveren et al. 1998).
During the RSG stage the stalled bubble can re-expand (D'Ercole 1992)
and catch up the moving star,
provided that there are no external sources of ionizing emission.
The high-velocity massive stars also could explode inside the large-scale
bubbles, but this happens only for stars whose mass $\geq 15-20 \,
{\rm M}_{\odot}$. In this case, a massive star before it exploded as a SN
becomes a Wolf-Rayet (WR) star (e.g. Vanbeveren et al. 1998),
whose fast wind blows up a new large-scale bubble
surrounded  by a dense shell. And again, the proper motion of the
WR star results in the significant offset of the SN explosion site
from the centre of the wind-driven bubble. E.g. for the
duration of the WR stage of $\simeq 2-3\times 10^5$ yr and the stellar
velocity of $30 \, {\rm km}\,{\rm s}^{-1}$ the SN explosion site will be
displaced from the centre of the bubble for about $6-9$ pc
(see Arnal 1992 for examples of non-central location of WR stars in
bubbles created by their winds). On the other hand, most of massive
stars are less massive than $15 M_{\odot}$, and therefore do not evolve
through the WR stage before the SN explosion. Thus, if a massive star
of mass $<15 M_{\odot}$ is fast enough to cross the MS bubble before it
exploded as a SN, the SN blast wave mainly interacts with the unperturbed
interstellar medium and the SN explosion site coincides with the
geometrical centre of the resulting (middle-aged) SNR. In this case the
shell of the SNR could appear as an incomplete circle, that is due
to the presence of a low-density tunnel created by the stellar wind
behind the moving star (see Brighenti \& D'Ercole 1994).

It follows from the above discussion that though many of SNRs are produced
by SN explosions outside of large-scale wind-driven bubbles (and
therefore their structure could be discribed in the framework of the
standard model based on the Sedov-Taylor solution), there are could
exist SNRs whose origin is connected with off-centred cavity SN
explosions\footnote{Though the knowledge of the fraction of these SNRs is
very important for statistical studies of NS/SNR associations, we
are now not in a position to quantify it. Two main difficulties
on the way to do this are the absence of
self-consisting evolutionary models for rotating stars (the evolutionary
paths of rotating stars differ from those of non-rotating ones) and
wind-driven bubbles (taking into account the heat conduction and
magnetic effects). These difficulties along with uncertainties in the
velocity distribution and initial mass function of massive stars do not
allow to solve the problem properly.}.
Therefore the high transverse velocities inferred for a number of NSs
through their association with SNRs could be reduced (see next Sect.).
The velocity
reduction could be large enough for NSs associated with middle-aged and old
SNRs, i.e. the SNRs whose origin could
be connected with the interaction of SN blast waves with
large-scale structures created by the fast stellar wind, and where the SN
explosion sites could be significantly offset from the centres of the
wind-driven bubbles.
But the velocity reduction should be less considerable in the case of
young ($< 10^3$ years) SNRs, whose appearance is mostly determined by
the interaction of SN blast waves with circumstellar (i.e. small-scale)
structures created during the late (RSG and WR) evolutionary stages of SN
progenitor stars (see Sect. 3). These stages are significantly
shorter than the MS stage and only sufficiently fast
stars have time to became significantly displaced from the centres of
associated circumstellar structures. Therefore, the geometrical centres of
young SNRs better correspond to the SN explosion sites, that explains the
quite symmetric appearance of these SNRs (a marked exception is the
Kepler's SNR, whose asymmetric shell is due to the very fast motion
of the SN progenitor star (Bandiera 1987)). But even the slow motion of
the SN progenitor star results in an appreciable asymmetry of
circumstellar structures (e.g. the mass distribution over the nearly
spherical circumstellar shell becomes inhomogeneous), that, for instance,
results in the asymmetric expansion of young SNRs. A good example of such
a young SNR is the Cas\,A. A compact X-ray
source was recently discovered near the geometrical centre of this SNR
(Tananbaum 1999). The implied transverse velocity of the compact
source derived
through the various determinations of the expansion centre of Cas\,A (e.g.
van den Berg \& Kamper 1983, Reed et al. 1995) ranges from $50$ to $1000
\, {\rm km}\,{\rm s}^{-1}$ (Pavlov et al. 2000).
We suggest that the separation of the SN explosion site (and the compact
source) from the geometrical centre of Cas\,A could be caused to a
large extent by the
proper motion of the SN progenitor star. This motion (with the velocity
less than the velocity of the RSG wind) will result in the
deviation from spherical (or axial) symmetry of the circumstellar matter,
that in its turn could be responsible for the observed (see Vink et al.
1998, and references therein) expansion asymmetry of Cas\,A.

\section{Discussion}

In Sect. 2 we showed that the large displacement of a NS
from the geometrical centre of the associated SNR does
not inevitably mean that this NS is moving with high transverse
velocity. This
could have an important impact on the understanding of the origin of the
phenomenon of anomalous X-ray pulsars and soft
gamma-ray repeaters (SGRs) since the high implied velocities of some of
these objects were interpreted as a sign that they represent
a high-velocity ($\sim 1000 \, {\rm km}\,{\rm s}^{-1}$) population of NSs
(e.g. Thompson \& Duncan 1995; Marsden et al. 1999). It seems
that the recent association of two of
known SGRs with clusters of massive stars (Fuchs et al. 1999, Vrba et
al. 2000) should reduce the acuteness of the problem of high implied
velocities of these objects, but the large angular offset of the
SGR\,0525-66 from the centre of the nearly spherical (in X-rays)
SNR N\,49 in LMC (e.g. Rothschild, Kulkarni \& Lingenfelter 1994)
still continues to raise doubts in the
association of these two objects (e.g. Kaspi 2000, Kaplan et al. 2001).
Note also that the high transverse velocities derived by Frail,
Goss \& Whiteoak (1994) for pulsars associated with SNRs
were used to put forward a number of quite strong suggestions,
e.g. that "SNRs are produced preferentially by the (SN) explosions that
yield fast kicks" (Cordes \& Chernoff 1998).

The high implied transverse velocities of NSs are sometimes
used to discard the possible NS/SNR associations. E.g. Stappers,
Gaensler \& Johnston (1999) suggested that the lack of a
pulsar wind radio nebula around the PSR B\,1610-50 means that the
maximum space velocity $v_{\rm p}$ of this pulsar is $450 \,
(d/5 \,{\rm kpc}) \, {\rm km}\,{\rm s}^{-1}$, where
$d$ is the distance to the pulsar, and therefore it could not be
associated with the nearby SNR Kes\,32 since this association implies
the transverse velocity of the pulsar of $\simeq 2000 \,
{\rm km}\,{\rm s}^{-1}$ (Caraveo 1993).
The implied transverse velocity, however, could be reduced two times simply
due to the possible off-centred SN explosion, and once again two or even
more times if the braking index of the pulsar is similar respectively to
that of the PSR B\,0540-69 (Boyd et al. 1995) or the Vela pulsar
(Lyne et al. 1996).

The association of PSR B\,1610-50 with SNR Kes\,32 was also recently
questioned by Pivovaroff, Kaspi \& Gotthelf (2000). They used
the non-detection of an X-ray nebula around the PSR B\,1610-50 to
estimate $v_{\rm p}$ to be less than $170 \, (d/7.3\,{\rm kpc})^2 \,
(n/1\,{\rm cm}^{-3} )^{-1/2} \, {\rm km}\,{\rm s}^{-1}$.
This estimate was derived under the assumption
that the wind of the PSR B\,1610-50 has
the same characteristics as that of the Crab pulsar,
and for the number density of
the ambient medium $n=1 \, {\rm cm}^{-3}$.
One can, however, show that reasonable
variations of the assumed parameters allow to
increase the estimated velocity of
the pulsar. E.g. for $n\leq 10^{-2} \, {\rm cm}^{-3}$
and $d=5$ kpc (Stappers et al. 1999),
one has $v_{\rm p} \leq 780 \, {\rm km}\,{\rm s}^{-1}$.

The high transverse velocity was also inferred for the pulsar PSR
B\,1757-24, which
lies well outside the shell of the SNR G\,5.4-1.2
(e.g. Caswell et al. 1987). The
physical association of these two objects was
firmly established after the discovery
(Frail \& Kulkarni 1991; see also Manchester et al. 1991) of a tail of radio
emission connecting the pulsar with the SNR. The pulsar PSR B\,1757-24 is,
however, more interesting in that that its proper motion vector
does not point away from the geometrical centre of the nearly circular
shell of the remnant (the radius of which is about 16 arcmin),
but misses it by nearly 5 arcmin
(Frail, Kassim \& Weiler 1994). To explain this inconsistency, Frail et al.
(1994) suggested that the SN exploded in an exponentially
stratified medium and used the Kompaneets (1960) solution to fit the
shape of the remnant. This allowed them to
put the possible SN explosion site closer to the present
position of the pulsar, that reduces the implied velocity of the pulsar
to the value between 1300 and 1700 ${\rm km}\,{\rm s}^{-1}$. We propose an
alternative explanation and suggest that the SNR G\,5.4-1.2 is the result
of the off-centred SN explosion in the pre-existing
wind-driven bubble surrounded by a massive  shell (Gvaramadze,
in preparation)\footnote{The mass of the shell
is a very important parameter since it determines the evolution of the
SNR. If the mass of the shell is larger than about 50 times the mass
of the SN ejecta (e.g. Franco et al. 1991), the SN blast wave merges with
the shell and evolves into a momentum-conserving stage (i.e. it skips the
Sedov-Taylor stage). In this case, even a young NS moving with a moderate
velocity ($\geq 200 \, {\rm km}\,{\rm s}^{-1}$) is able to
overrun the SNR's shell (cf. Gaensler \& Johnston 1995a),
provided that it was born not far from the edge of the wind-driven
bubble.}.
This suggestion allows to reduce considerably the transverse velocity of the
pulsar and naturally explains why the tail behind the pulsar does not
point back to the centre of the remnant.
An indirect support of our suggestion comes from the recent observations
of the radio nebula surrounding the pulsar (Gaensler \& Frail 2000). These
observations put an upper limit on the pulsar proper motion, which turns
out to be much smaller than that expected if the
pulsar velocity is indeed in the range derived by Frail et al. (1994).
Assuming that the pulsar was born in the geometrical centre of the
associated SNR, Gaensler \& Frail (2000) argued that the
true age of the pulsar should be more than 10 times larger than the
characteristic age (cf. Istomin 1994). The off-centred
cavity SN explosion provides another possible explanation for the low
value of the pulsar proper motion.

Let us discuss the third criterion for the evaluation of NS/SNR associations
suggested by Kaspi (1996).
There are three factors which could affect the estimates
of the characteristic age of a NS (e.g. Camilo 1996).
First, the braking index could
be different from that follows
from the simplest spindown models (see e.g. Lyne
et al. 1993, Kaspi et al. 1997, Marshall et al. 1998).  Second, the NS
could be born with the large initial spin period (e.g. Spruit \& Phinney
1998), and therefore the true age could be much smaller than the
characteristic one.
Third, the spindown torque (as well as the braking index)
could be a function of time. E.g. if the spindown of a NS is
due to the magnetic dipole radiation, than the
secular increase of the magnetic moment of the NS results
in the increase of the braking torque (e.g. Blandford \& Romani 1988).
The spindown rate of a NS could be also enhanced due to the
interaction of its magnetosphere with the dense ambient medium (Istomin 1994,
Yusifov et al. 1995, Gvaramadze 1999c, 2001, Menou, Perna \& Hernquist
2001). In both cases the true age of the NS could be much larger than the
characteristic age derived from the present value of the spin period
derivative.  These arguments were used to reconcile the ages of the pulsar
PSR B\,1509-58 and the associated SNR MSH\,15-52 (Blandford \& Romani
1988, Gvaramadze 1999c, 2001), or to show that the implied transverse
velocity of the pulsar PSR B\,1757-24 could be reduced (Istomin 1994).

In Gvaramadze (1999c, 2001) we suggested that the high spin-down rate of the
pulsar PSR B\,1509-58 is inherent only for a relatively short period of
its present spin history, and that the enhanced braking torque is caused by
the interaction of the pulsar's magnetosphere with the material of a
dense circumstellar clump created during the late stages of evolution of
the SN progenitor star.  The origin of dense circumstellar clumps could be
explained in the framework of the three-wind model (e.g. Garsia-Segura et
al. 1996). The fast (WR) wind sweeps the slow (RSG) wind and creates a
low-density cavity surrounded by a shell of swept-up circumstellar matter.
This shell expands with the nearly constant velocity $v_{\rm sh} \simeq
(\dot{M}_{\rm WR} v_{\rm WR} ^2 v_{\rm RSG} /3\dot{M}_{\rm RSG})^{1/3}$,
where $\dot{M}_{\rm WR}, \dot{M}_{\rm RSG}$ and $v_{\rm WR}, v_{\rm RSG}$
are, correspondingly, the mass-loss rates and wind velocities during the
WR and RSG stages (e.g. Dyson 1981), until it catches up the shell
separating the RSG wind from the MS bubble (the characteristic radius of
this shell is a few pc; the high-pressure gas of the MS bubble interior
hinders the free expansion of the RSG wind (e.g. Chevalier \& Emmering
1989, D'Ercole 1992)). The interaction of two
circumstellar shells results in the Rayleigh-Taylor and other dynamical
instabilities, whose development is accompanied by the formation of dense
clumps moving with radial velocities of $v_{\rm cl} \simeq
v_{\rm sh}$ (Garsia-Segura et al. 1996). For parameters typical
for RSG and WR winds, one has $v_{\rm
cl} \simeq 100-200 \, {\rm km}\,{\rm s}^{-1}$. The radial velocity of
quasi-stationary flocculy in Cas\,A (whose origin could be attributed to
the processes discussed above; e.g. Garsia-Segura et al. 1996) ranges from
$\simeq 80$ to $\simeq 400 \,{\rm km}\,{\rm s}^{-1}$. The dense clumps
could originate much closer to the SN progenitor star due to the stellar
wind acceleration during the transition from the RSG to the WR stage
(Brighenti \& D'Ercole 1997).
After the SN exploded, the SN blast wave propagates through the tenuous
interclump medium, leaving behind the dense clumps embedded in the hot
shocked interclump gas (the filling factor of the clumps is
small (e.g. Gvaramadze 2001) and therefore they do not affect considerably
the dynamics of the SN blast wave).
The gradual evaporation of the dense material of radially
moving clumps results in the origin of an expanding nebula
of thermal X-ray emission, which marks the SN explosion site.

It is clear from the above discussion that the nebulae of thermal X-ray
emission should exist only in those SNRs, whose origin is connected with
explosions of massive stars with zero age main sequence mass $\geq 15-20
M_{\odot}$ (only in these cases one can expect that the (clumpy)
circumstellar material will survive the passage of the SN blast wave).
It is clear also that
the motion of the SN progenitor star could result in a significant offset
of the compact region of dense circumstellar matter from the centre of the
MS bubble (the RSG and WR stages are about 10-20
times shorter than the MS stage), and correspondingly in the
offset of the nebula of thermal X-ray emission from the geometrical
centre of the associated middle-aged SNR.
The possible detection of such nebulae will provide
the direct observational test for our proposal, and could be used for the
re-estimation of transverse velocities of the already known NSs, or for
the search of new stellar remnants possibly associated with these SNRs.

To find a crude order of magnitude estimate of the luminocity of the
nebula of thermal X-ray emission,
we assume that all the X-ray emitting interclump material (including
the gas already evaporated from the clumps) is at the same
temperature between
$10^7$ and $10^8$ K, and that this material is uniformly dispersed over a
sphere of radius $R=R_0 + v_{\rm cl} t_{\rm SNR}$, where $R_0 \, (\simeq
1-2$ pc) is the radius of the region occupied by dence clumps at the
moment of SN explosion, $t_{\rm SNR}$ is the age of the SNR, then one has
$L_{\rm x} \simeq 1.2\times 10^{33} n^2 R_{\rm pc} ^3 \, {\rm ergs}\,{\rm
s}^{-1}$, where $n$ is the number density of the emitting gas, $R_{\rm pc}
=R/1\,{\rm pc}$ (Gorenstein \& Tucker 1976). For $R_0 =2$ pc,
$v_{\rm cl} =100 \, {\rm km}\,{\rm s}^{-1}$ and $t_{\rm SNR} =10^4$ years,
and assuming that the mass of the emitting gas is
$\simeq 5M_{\odot}$ (i.e. about a half of the mass lost by a
$15M_{\odot}$ star during the RSG stage), one has $R\simeq 3$ pc and
$L_{\rm x} \simeq 4\times 10^{34} \, {\rm ergs}\,{\rm s}^{-1}$.
The similar estimates were used by Gvaramadze (1999a)
to show that the nebula of hard X-ray emission found by Willmore et al.
(1992) around the Vela pulsar could be the dense material lost by
the SN progenitor star in the form of the RSG wind and heated to the
observed temperature after the SN exploded. The more detailed analysis of
this problem constitutes a part of a project underway to study the origin of
mixed-morphology SNRs (Rho \& Petre 1998) and will be published elsewhere
(for a different point of view see e.g. White \& Long 1991 and Petruk 2000).

In conclusion we note that in Gvaramadze (1999c) we interpreted a bright
X-ray spot (which nearly coincides with the error box for the
SGR\,0525-66; Rothschild et al. 1994) on the periphery of the SNR N\,49
as an X-ray nebula marking the SN explosion site and suggested that the
large implied transverse velocity of the NS associated with the SGR
could be reduced about ten times. In our analysis we assumed that the spot is
a thermal feature (cf. Dickel et al. 1995) and that the radius of the spot
is about $5'' -10''$ (i.e. $\simeq 1-2$ pc; see Rothschild et al. 1994
and Dickel et al. 1995). We found that to explain the X-ray luminocity of
the spot of $\simeq 10^{36} \,{\rm ergs}\,{\rm s}^{-1}$ (Rothschild
et al. 1994), the mass of the X-ray emitting gas should be in a
range $4-10 M_{\odot}$ (i.e. a reasonable value, provided that the zero
age main sequence mass of the SN progenitor star was $\geq 15M_{\odot}$).
However, recent high-resolution
Chandra X-ray Observatory observations of the X-ray spot in N\,49 (Kaplan
et al. 2001 and references therein) showed that this source is pointlike
and could be considered as the X-ray counterpart of SGR\,0525-66.
Though this result discards our interpretation of the spot as an X-ray
nebula, we believe that the large angular displacement of SGR\,0525-66
from the centre of N\,49 is due to the effect discussed in this Letter.

\section{Summary}

A cavity SN explosion of a moving massive star could result in a
significant offset
of the NS birth-place from the centre of the nearly spherical
middle-aged SNR. Therefore: a) the
high transverse velocities inferred for a number
of NSs (e.g.  PSR B\,1610-50, PSR B\,1757-24, SGR\,0525-66)
through their association with SNRs could be reduced;
b) the proper motion vector of the NS should not necessarily point
away from the geometrical centre of the associated SNR.  These
two facts allow to enlarge the circle of possible NS/SNR associations
and should be taken into account in evaluating of their reliability.
The birth-place of the NS could be marked by a (compact) nebula of thermal
X-ray emission. The discovery of such nebulae in middle-aged SNRs could be
used for the re-estimation of transverse velocities of the already
known NSs, or for the search of new stellar
remnants possibly associated with these SNRs.
\vskip 5mm

{\bf Acknowledgements}. I am grateful to N.D'Amico and A.D'Ercole for their
interest to this work and to the anonymous referee for her/his questions
and comments allowing me to clarify some points discussed in the Letter.


\begin{thebibliography}{}
%
\bibitem{} Allakhverdiyev A.O., G\"ok F., H\"useyinov O.H., Tuncer E.,
\"Ogelman H.B., 1995, in Alpar M.A., Kizilo\v{g}lu \"U., van Paradijs J., eds,
The Lives of the Neutron Stars. Kluwer, Dordrecht, p. 43
\bibitem{} Arnal E.M., 1992, A\&A, 254, 305
\bibitem{} Avedisova V.S., 1972, Sov. Astron., 15, 708
\bibitem{} Bandiera R., 1987, ApJ, 319, 885
\bibitem{} Blandford R.D., Romani R.W., 1988, MNRAS, 234, 57p
\bibitem{} Borkowski K.J., Szymkowiak A.E., Blondin J.M., Sarazin C.L., 1996,
ApJ, 466, 866
\bibitem{} Boyd P.T. et al., 1995, ApJ, 448, 365
\bibitem{} Brighenti F., D'Ercole A., 1994, MNRAS, 270, 65p
\bibitem{} Brighenti F., D'Ercole A., 1997, MNRAS, 285, 387
\bibitem{} Camilo F., 1996, in Johnston S., Walker M.A., Bailes M.,
eds, Pulsars: Problems and Progress. ASP, San Francisco, p. 39
\bibitem{} Caraveo P.A., 1993, ApJ, 415, L111
\bibitem{} Caraveo P.A., 1995, in Alpar M.A.,
Kizilo\v{g}lu \"U., van Paradijs J.,
eds, The Lives of the Neutron Stars. Kluwer, Dordrecht, p. 39
\bibitem{} Castor J., McCray R., Weaver R., 1975, ApJ, 200, L107
\bibitem{} Caswell J.L., Kesteven M.J., Komesaroff M.M., Haynes R.F.,
Milne D.K., Stewart R.T., Wilson S.G., 1987, MNRAS, 225, 329
\bibitem{} Chevalier R.A., Emmering R.T., 1989, ApJ, 342, L75
\bibitem{} Chevalier R.A., Liang E.P., 1989, ApJ, 344, 332
\bibitem{} Ciotti L., D'Ercole A., 1989, A\&A, 215, 347
\bibitem{} Cordes J.M, Chernoff D.F., 1998, ApJ, 505, 315
\bibitem{} Dickel J.R., et al., 1995, ApJ, 448, 623
\bibitem{} Dyson J.E., 1975, Ap\&SS, 35, 299
\bibitem{} Dyson J.E., 1981, in Kahn F.D., ed., Investigating the
Universe. Reidel, Dordrecht, p. 125
\bibitem{} Dyson J.E., de Vries J., 1972, A\&A, 20, 223
\bibitem{} D'Ercole A., 1992, MNRAS, 255, 572
\bibitem{} Fabian A.C., Brinkmann W., Stewart G.C.,
1983, in Danziger J., Gorenstein
P., eds, Supernova Remnants and Their X-Ray Emission.
Reidel, Dordrecht, p. 119
\bibitem{} Frail D.A., 1998, in Buccheri R.,
van Paradijs J., Alpar M.A., eds, The
Many Faces of Neutron Stars. Kluwer, Boston, p. 179
\bibitem{} Frail D.A., Kulkarni S.R., 1991, Nat, 352, 785
\bibitem{} Frail D.A., Goss W.M., Whiteoak J.B.Z., 1994, ApJ, 437, 781
\bibitem{} Frail D.A., Kassim N.E., Weiler K.W., 1994, AJ, 107, 1120
\bibitem{} Franco J., Tenorio-Tagle G., Bodenheimer P., Rozyczka M., 1991,
PASP, 103, 803
\bibitem{} Fuchs Y., Mirabel F., Chaty S., Claret A.,
Cesarsky C.J., Cesarsky D.A., 1999, A\&A, 350, 891
\bibitem{} Gaensler B.M., Frail D.A., 2000, Nat, 406, 158
\bibitem{} Gaensler B.M., Johnston S., 1995a, MNRAS, 275, L73
\bibitem{} Gaensler B.M., Johnston S., 1995b, MNRAS, 277, 1243
\bibitem{} Garsia-Segura G., Langer N., Mac Low M.-M., 1996, A\&A, 316,
133
\bibitem{} Gorenstein P., Tucker W.H., 1976, ARAA, 14, 373
\bibitem{} Gvaramadze V.V., 1999a, A\&A, 352, 712
\bibitem{} Gvaramadze V.V., 1999b, Odessa Astron. Publ., 12, 117 (or
astro-ph/9912512)
\bibitem{} Gvaramadze V.V., 1999c, in Kardashev N.S., Dagkesamansky R.D.,
Kovalev Yu.A., eds, Astrophysics on the Boundary of Centuries.
ASC, Moscow, p. 163 (in Russian)
\bibitem{} Gvaramadze V.V., 2001, astro-ph/0102431
\bibitem{} Helfand D.J., 1998, Mem. Soc. Astron. It., 69, 791
\bibitem{} Istomin Ya.N., 1994, A\&A, 283, 85
\bibitem{} Kaplan D.L. et al., 2001, astro-ph/0103179
\bibitem{} Kaspi V.M., 1996, in Johnston S., Walker M.A., Bailes M., eds,
Pulsars: Problems and Progress. ASP, San Francisco, p. 375
\bibitem{} Kaspi V.M., 1998, in Shibazaki N., Kawai N., Shibata S.,
Kifune T., eds, Neutron Stars and Pulsars: Thirty Years after the
Discovery. Universal Academy Press, Tokyo, p. 401
\bibitem{} Kaspi V.M., 2000, in Kramer M., Wex N., Wielebinski R., eds, Pulsar
Astronomy -- 2000 and Beyond. ASP, San Francisco, p. 485
\bibitem{} Kaspi V.M., Bailes M., Manchester R.N., Stappers B.W., Sandhu
J.S., Navarro J., 1997, ApJ, 485, 820
\bibitem{} Kompaneets A., 1960, Sov. Phys. Dokl., 5, 46
\bibitem{} Lyne A.G., Pritchard R.S., Graham-Smith F., Camilo F., 1996,
Nat, 381, 497
\bibitem{} Lorimer D.R., Lyne A.G., Camilo F., 1998, A\&A, 331, 1002
\bibitem{} McCray R., 1983, Highlights Astr., 6, 565
\bibitem{} McCray R., 1993, ARA\&A, 31, 175
\bibitem{} McKee C.F., 1988, in: Supernova Remnants and the Interstellar
Medium, eds. R.S.Roger, T.L.Landeker, Cambridge: Cambridge Univ. Press, p. 205
\bibitem{} McKee C.F., Van Buren D., Lazareff R., 1984, ApJ, 278, L115
\bibitem{} Manchester R.N., Kaspi V.M., Johnston S., Lyne A.G.,
D'Amico N., 1991, MNRAS, 253, 7p
\bibitem{} Marsden D., Lingenfelter R.E., Rothschild R.E., Higdon J.C., 1999,
astro-ph/9912207
\bibitem{} Marshall F.E., Gotthelf E.V., Zhang W., Middleditch J., Wang
Q.D., 1998, ApJ, 499, L179
\bibitem{} Menou K., Perna R., Hernquist L., 2001, astro-ph/0103326
\bibitem{} Mereghetti S., 1998, Mem. Soc. Astr. It., 69, 819
\bibitem{} Mereghetti S., 1999, astro-ph/9911252
\bibitem{} Pavlov G.G., Zavlin V.E., Aschenbach B., Tr\"umper J.,
Sanwal D., 2000, ApJ, 531, L53
\bibitem{} Petruk O, 2000, astro-ph/0006161
\bibitem{} Pivovaroff M.J., Kaspi V.M., Gotthelf E.V., 2000, ApJ, 528, 436
\bibitem{} Reed J.E., Hester J.J., Fabian A.C., Winkler P.F., 1995, ApJ,
440, 706
\bibitem{} Rho J., Petre R., 1998, ApJ, 503, L167
\bibitem{} Rothschild R.E., Kulkarni S.R., Lingenfelter R.E., 1994, Nat, 368, 432
\bibitem{} R\'{o}\.{z}yczka M., Tenorio-Tagle G., Franco J., Bodenheimer P., 1993,
MNRAS, 261, 674
\bibitem{} Shull P., Dayson J.E., Kahn F.D., West K.A., 1985, MNRAS, 212, 799
\bibitem{} Spruit H., Phinney E.S., 1998, Nat, 393, 139
\bibitem{} Stappers B.W., Gaensler B.M., Johnston S., 1999, MNRAS, 308, 609
\bibitem{} Steigman G., Strittmatter P.A., Williams R.E., 1975, ApJ, 198, 575
\bibitem{} Tammann G.A., L\"offler W., Schr\"oder A., 1994, ApJS, 92, 487
\bibitem{} Tananbaum H., 1999, IAUC 7246
\bibitem{} Thompson C., Duncan R.C., 1995, MNRAS, 275, 255
\bibitem{} Vanbeveren D., De Loore C., Van Rensbergen W., 1998, A\&AR, 9,
63
\bibitem{} van den Berg S., Kamper K.W., 1983, ApJ, 268, 129
\bibitem{} van den Berg S., Tammann G.A., 1991, ARA\&A, 1991, 29, 363
\bibitem{} Vink J., Bloemen H., Kaastra J.S., Bleeker J.A.M., 1998, A\&A,
339, 201
\bibitem{} Vrba F.J., Henden A.A., Luginbuhl C.B., Guetter H.H., Hartmann D.H.,
Klose S., 2000, ApJ, 533, L17
\bibitem{} Weaver R., McCray R., Castor J., Shapiro P., Moore R., 1977,
ApJ, 218, 377
\bibitem{} White R.L., Long K.S., 1991, ApJ, 373, 543
\bibitem{} Willmore A.P., Eyles C.J., Skinner G.K., Watt M.P., 1992,
MNRAS, 254, 139
\bibitem{} Yusifov I.M., Alpar M.A., G\"ok F., H\"useyinov O.H., 1995,
in Alpar M.A., Kizilo\v{g}lu \"U., van Paradijs J., eds,
The Lives of the Neutron Stars. Kluwer, Dordrecht, p. 201

\end{thebibliography}
\end{document}